# Blockchain-Assisted Spectrum Trading between Elastic Virtual Optical Networks

Shifeng Ding, Kevin X. Pan, Sanjay K. Bose, Qiong Zhang, and Gangxiang Shen

***Abstract*— In communication networks, network virtualization can usually provide better capacity utilization and quality of service (QoS) than what can be achieved otherwise. However, conventional resource allocation for virtualized networks would still follow a fixed pattern based on the predicted capacity needs of the users, even though, in reality, the actual traffic demand of a user will always tend to fluctuate. The mismatch between the fixed capacity allocation and the actual fluctuating traffic would lead to degradation of provisioned network services and inefficiency in the assigned network capacity. To overcome this, we propose a new spectrum trading (ST) scheme between virtual optical networks (VONs) in the context of an elastic optical network (EON). The key idea here is to allow different VONs to trade their spectrum resources according to the actual capacity they need at different time instants. A VON with unused spectra can then trade away its unused spectra to other VONs that are short of spectrum resources at that time. In exchange, it is rewarded with a certain amount of credit for its contribution to the ST community, which it can then use later to get extra bandwidth, if needed. The trust-worthiness of the trading records between the VONs is ensured in a distributed fashion through a blockchain-assisted account book that is updated whenever a new trade occurs. For this, we develop a software-defined control plane to enable spectrum trading in an EON. The performance of the ST scheme is evaluated and compared with a scenario without such trading. Our results show that the proposed ST scheme is effective in improving the QoS of each VON and significantly improves the overall network capacity utilization.**

## I. Introduction

Internet traffic is rapidly increasing with the proliferation of Internet of Things (IoT) and bandwidth-intensive services such as high definition video and virtual reality (VR). According to Cisco, not only will the annual global IP traffic reach 4.8 ZB per year by 2022, but also that the busy-hour Internet traffic is growing more rapidly than average Internet traffic [1] which would put excess stress on network resources. This, in turn, greatly increases the pressure on the optical network providing the underlying infrastructure for the upper-layer services. To relieve this pressure, network virtualization is implemented to divide an optical network into multiple virtual optical networks (VONs) that share a common physical hardware [2]. Here, efficiently assigning capacity to each VON, through *Virtual Optical Network Embedding* (VONE) [3] would be important to get good overall capacity utilization of the optical network. Most existing studies on VONE assume fixed capacity requirements, which assign a fixed amount of bandwidth on each VON. However, in reality, actual traffic demand on a VON will always fluctuate over time, causing a mismatch between the assigned network resources and the actual capacity requirements. Specifically, when the traffic demand on a VON is low, the assigned capacity to the VON is overprovisioned and the network capacity is under-utilized. On the contrary, when the traffic demand on the VON is high, the assigned capacity to the VON may not be enough, degrading the quality of network services. To overcome this mismatch problem, some approaches have been developed for flexible VON resource assignment. Chen *et al.* [4][5] proposed a gaming model to motivate tenants to provide virtualized network services based on revenue and QoS incentives. In [6], Zhu *et al.* designed a network system for on-demand application-driven network slicing. These approaches lead to a more flexible mode of operation allowing the carrier to assign different amounts of bandwidth to users at different times.

In this paper, we propose a novel *Spectrum Trading* (ST) scheme in the context of an EON. This allows VON clients to trade their spectrum resources according to their actual capacity requirements. Clients with more capacity than what they currently need can provide their excess capacity to other clients that need more than what they currently have. The contribution of spectrum resources by each client to the ST community is in turn rewarded with a certain amount of credit which can be used to obtain resources from other clients, if needed in the future [7]. For a large-scale ST community with many VON clients, issues on the efficiency and security of such trading need to be carefully considered. These include

- How to implement the ST scheme among VONs so that the benefit of the whole trading community is maximized?
- How to prevent the selfishness of some clients and also ensure that all the clients can fairly get spectrum resources from the trading community whenever they need it?
- How to guarantee the security of trading records and prevent them from being maliciously tampered with?

This paper gives a detailed description of the ST framework proposed for this. Our main contributions here are:

- We propose a ST scheme that enables VON clients to trade their spectrum resources according to their actual traffic

This work was jointly supported by National Natural Science Foundation of China (NSFC) (61671313) and a Project Funded by the Priority Academic Program Development of Jiangsu Higher Education Institutions.

Shifeng Ding and Gangxiang Shen are with Soochow University; Kevin X. Pan is with Twitch.tv (Amazon subsidiary); Qiong Zhang is with Fujitsu Labs of America; Sanjay K. Bose is with IIT Guwahati.

demands. The proposed scheme can significantly improve the spectrum utilization of an EON.

- To guarantee the security of the ST scheme and prevent trading records from being maliciously tampered with, we develop a blockchain-based database to store all the trading data, which is commonly maintained by all the VON clients in the ST community.
- We evaluate the efficiency of the ST scheme. Numerical results show that it can significantly increase the total traffic carried in an EON and improve the QoS of each VON.
- Based on the proposed ST scheme, we also highlight some open issues that may be interesting to readers.

The proposed ST scheme differs significantly from the mechanism of dynamically re-allocating spectrum and reconfiguring VONs though both can improve network resource utilization. However, the latter is only applied between a network operator and a VON client. The latter needs to pay more when more network resources are assigned, but is not refunded when its assigned network capacity is not fully used. In contrast, the ST scheme achieves efficient network resource utilization by forming a mutually beneficial community, which does not need to request new resources from the network operator, thereby avoiding extra payments.

## II. Spectrum Trading Scheme

We introduce the ST scheme here describing the basic trading concept, credit definition, fairness maintenance, and trading pair selection.

### *A. Basic Trading Concept*

Based on network virtualization, a client acquires a VON from a network operator to deliver its traffic. The VON is made up of a set of virtual links connecting multiple virtual nodes. Each virtual node is embedded in a physical node of the operator's network and each virtual link is a capacity pipe with a certain amount of bandwidth provided by a lightpath established between a pair of physical nodes. The amount of capacity assigned to each VON is often decided by a forecasted traffic demand, predicted based on historic traffic demand data. In most cases, this capacity cannot be frequently modified (either in amount or in topology) once it is assigned, since it is based on a service contract between the VON client and the network operator.

However, the traffic demands on the VONs fluctuate over time, which may lead to inefficient network resource utilization. Moreover, the intensities of traffic demands on the VONs are often asynchronous which implies that while some VONs may have transiently high traffic demands requiring more network resources, other VONs may have low traffic demands and have unused spectrum resources at that time. This creates an opportunity in trading spectrum resources between VONs, which is referred to as the *Spectrum Trading* (ST) scheme. For this, we divide the lifecycle of VONs into multiple time slots (typically in a uniform fashion). In each time slot, VONs with high traffic demands can potentially use the unused capacities owned by VONs with low traffic demands. For efficient trading, it is important to have a reliable traffic prediction algorithm

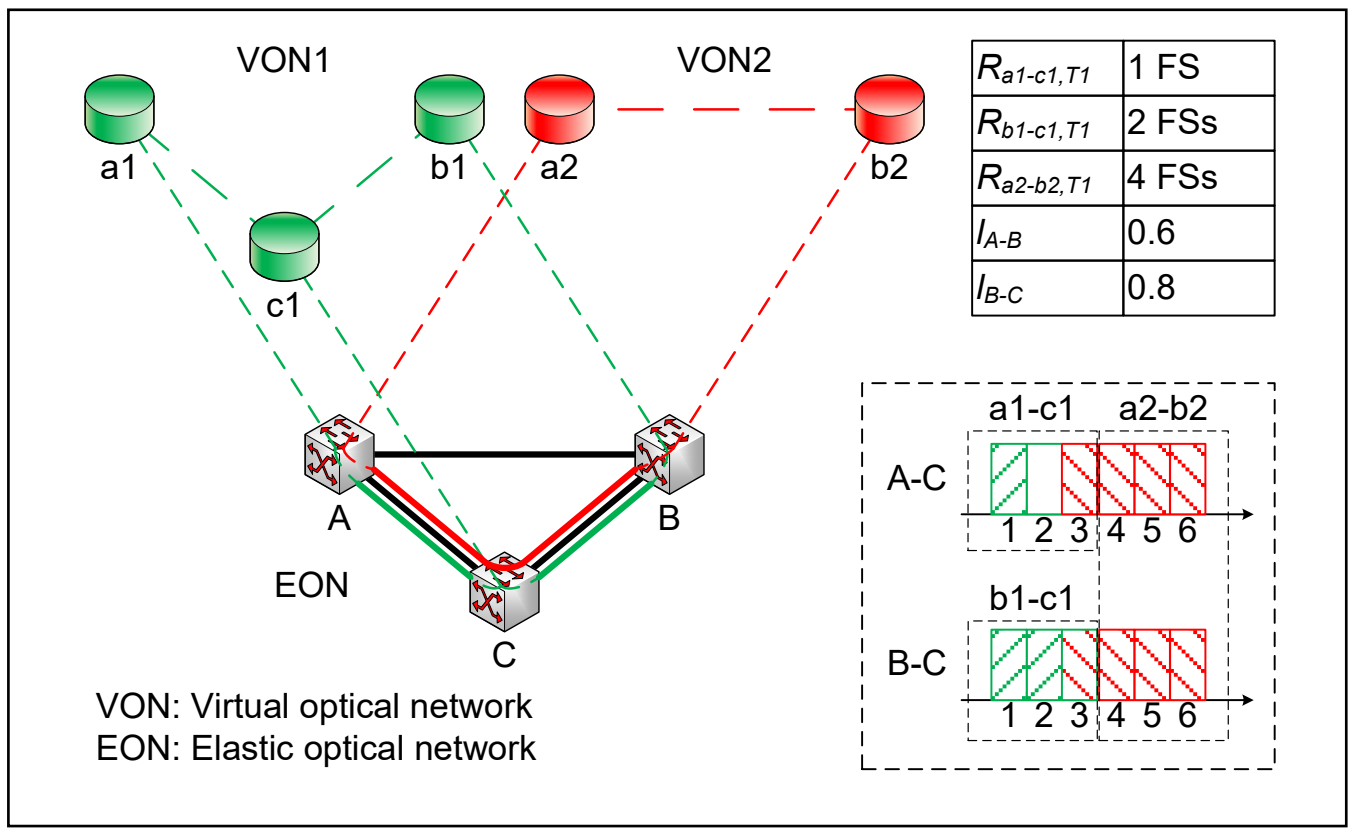

**Figure 1.** An example of spectrum trading.

[8][9] to predict the extra capacity that a VON may need and the unused capacity a VON may have in each time slot. Based on this prediction, the clients trade spectrum resources to improve the QoS of the VONs that have high traffic demands and maximize the overall traffic carried by the VONs.

Figure 1 uses an example to illustrate spectrum trading. Two virtual optical networks (i.e., VON1 and VON2) are embedded in a physical EON. Virtual links a1-c1 and b1-c1 of VON1 are mapped in lightpaths A-C and B-C respectively and virtual link a2-b2 of VON2 is mapped in lightpath A-C-B. Initially, all virtual links of the two VONs are assigned with 3 FSs as per the contracts between the VON clients and the network operator. In the next time slot (i.e., time slot T1), the capacity requirements of virtual links a1-c1, b1-c1, and a2-b2 are assumed to change to 1, 2, and 4 FSs, respectively. Without ST, 25% traffic on a2-b2 is blocked since its allocated capacity is not sufficient to satisfy the requirement. Meanwhile, two FSs of a1-c1 and one FS of b1-c1 are wasted since their capacities are overprovisioned. In this scenario, if we allow ST, a2-b2 can use the unused FSs of a1-c1 and b1-c1 on physical route A-C-B to accommodate its extra traffic (see FS occupation in Fig.1). This ensures that all the user traffic can be accommodated and less capacity is wasted overall in this time slot.

### *B. Credit Definition and Fairness Maintenance*

The ST scheme can help resource strapped clients to accommodate their traffic demands and improve the overall utilization of spectrum resources. However, this scheme may lead to a fairness issue if some selfish VON client constantly asks for spectrum resources from others, but rarely contributes its resources to others. Therefore, a fairness maintenance approach is required. We quantitatively define a trading credit to measure the spectrum resource contribution made by each VON. Also, to prevent a VON from being selfish and always asking for spectrum resources while not providing spectrum resources to others, we set a rule for VONs to use the unused spectrum resources in the ST community.

In each time slot, the credit of contribution made by a VON is defined as $C = \sum_{i \in \boldsymbol{L}} S_i * l_i$, where $\boldsymbol{L}$ is the set of physical links traversed by a VON, $S_i$ is the number of frequency slots (FSs) contributed by the VON on physical link $i$, and $l_i$ is the normalized physical length of link $i$. Based on the example in

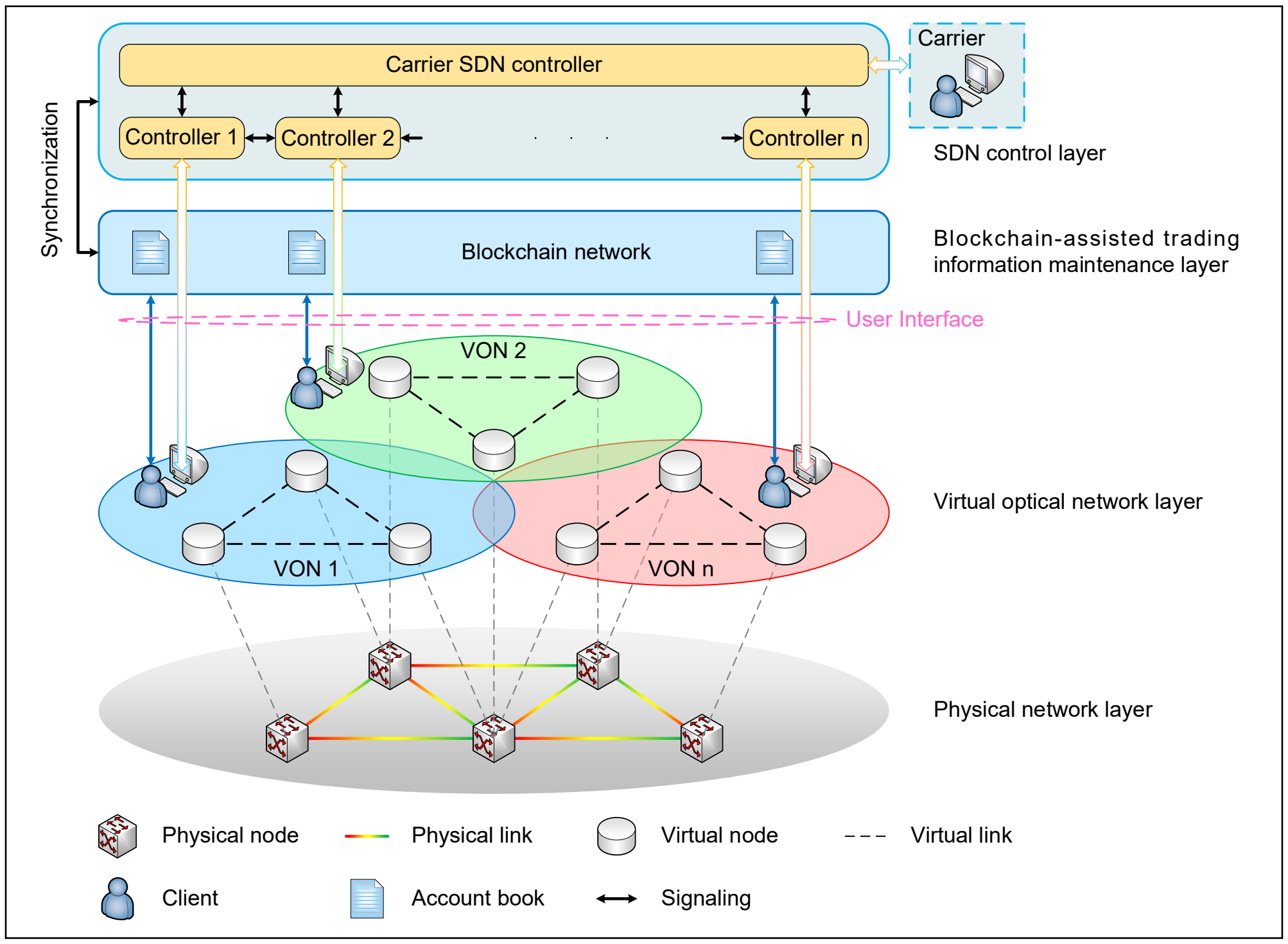

**Figure 2.** Architecture of the spectrum trading system.

Fig. 1, VON1 provides 1 FS to VON2 on physical route A-C-B. Then we can calculate the credit of VON1 in the current time slot as $C$=1×0.6+1×0.8=1.4 based on the normalized physical lengths of the links given in Fig. 1.

A cumulative credit is maintained for each VON, which is added to or subtracted from depending on whether a VON is providing to or using spectrum resources from the community. If a VON contributes spectrum resources to the community, its cumulative credit increases; otherwise, its cumulative credit decreases. In the example of Fig. 1, the cumulative credit of VON1 is increased with a credit of 1.4 while that of VON2 is reduced with a credit of 1.4.

To prevent a selfish VON from constantly using spectrum resources of the community, while rarely contributing its own resources to the community, we set a *forbidden threshold* for each VON's cumulative credit. Whenever the cumulative credit is smaller than this threshold, we prohibit the VON from using the spectrum resources of the community. Only when the cumulative credit becomes larger than the threshold (i.e., after contributing to the community), can it resume requesting spectrum resources from the community. Clearly, a proper threshold value would ensure an efficient ST system.

### *C. Trading Pair Selection*

To trade spectrum resources efficiently between different VONs, it is important to select trading pairs properly. Next we discuss the strategies for selecting the trading pairs. We first define different client roles, and then based on this, we elaborate on the principles of trading pair selection.

VON clients in the ST system have different roles:

- **Requesting client (RC):** A VON client that is short of spectrum resources and seeks unused spectrum resources from others to accommodate its traffic.
- **Candidate client (CC):** A VON client that has unused spectrum resources and can trade with a RC. Note that if the client's unused spectrum resources are not eligible to being used by the RC, it should not be considered as a CC even though it has unused spectrum resources. To judge whether a CC's spectrum resources are eligible to support the requirement of the RC, we need to consider the following two aspects. First, spatially we need to see if the RC and the CC overlap on any common physical link. Only if they share any physical link, can they have opportunities of trading spectrum resources. Moreover, on any shared physical link, we need to check their spectrum neighboring relationship. We may require the spectra of the CC and the RC to be neighboring on the physical link if the transponders used for setting up lightpaths do not support sub-band virtual concatenation (VCAT) [10].
- **Target client (TC):** A CC that is finally selected to offer unused spectrum resources to a RC is called TC.

Trading pair selection is a process to choose suitable TCs from the CCs to obtain unused spectrum resources required by RCs. A RC can use spectrum resources owned by multiple TCs simultaneously as long as the resources that they provide can jointly meet the RC's requirement. Similarly, a TC can also provide its spectrum resources to multiple RCs as long as these allocations are distinct and do not overlap.

For efficient trading pair selection, we need to check the cumulative credit of each VON client. A CC with the lowest cumulative credit should be selected first because it has used a large amount of spectrum resource of others, and it would be required for this CC to contribute to the community. Likewise, a CC should provide its unused spectrum resources to a RC with the highest cumulative credit first as the latter has contributed most to the community. These ensure the fairness of the trading mechanism.

## III. Software Defined and Blockchain-Assisted Spectrum Trading System

The ST system is an ecosystem collaboratively formed by the network operator and VON clients, who play different roles in the system. Specifically, the network operator owns the underlying physical resources and runs *Management and Orchestration* (MANO) functions. The clients purchase network resources from the operator in the form of VONs. After VON resources are assigned, each client has full control of its own resources over its lifetime. A VON client can monitor actual traffic demands on its virtual links and predict whether the links will need more or less bandwidth in the future. Based on this prediction, it decides whether spectrum trading is required and which VONs it will trade with. This process is carried out in a distributed fashion involving all the VONs that join the ST community. The trading data is stored in the form of a blockchain, which is collaboratively maintained by all the involved VON clients, not by the network operator. However, for each agreed trading, the carrier's network central controller physically configures the bandwidth in a suitable fashion for the VONs involved in the trading.

To implement this system, a powerful and robust control and management plane is necessary. Here we employ the software defined networking (SDN) technique for network control and management in addition to the blockchain-based technique for maintaining the spectrum trading information. As shown in Fig. 2, the ST system consists of four layers, including the physical network layer, the virtual optical network (VON) layer, the SDN control layer, and the blockchain-assisted trading information maintenance layer. These are described in detail next.

### *A. Physical Network Layer*

The physical network layer essentially corresponds to an EON, which contains optical cross-connects (OXCs) and fiber links interconnecting OXCs. In each OXC, there are add/drop ports connecting to optical transponders. The physical network layer provides the actual network capacity, in which multiple VONs are embedded.

### *B. Virtual Optical Network Layer*

In the VON layer, multiple VONs with diverse requirements are created for different clients and are independently controlled by these clients. Each VON consists of several virtual nodes and virtual links. Each virtual node is a virtualized network element with computing, forwarding, and storage resources. Each virtual link is a lightpath established in the physical network layer. Here provisioning bandwidth to a virtual link means that the transponders that initiate this bandwidth are also assigned for use and the VON user does not need to know exactly which transponders provide this bandwidth. This operational mode makes the assignment of bandwidth resources transparent to the associated transponders. In addition, all the VONs are commonly established on the underlying physical infrastructure, offering isolated services. To ensure that the cross-impairments among different VONs are within an acceptable range of signal transmission quality, a guard-band may be required between neighboring lightpaths of different VONs.

### *C. SDN Control Layer*

A SDN architecture is employed in our ST system. As shown in Fig. 2, the SDN controller layer comprises a carrier SDN controller and a set of distributed client SDN controllers, each of which has an exclusive control over a VON. The client SDN controllers receive instructions or requirements from their clients and relay them to the carrier SDN controller. They also extract an abstract view of the VON back to the clients [11]. The functions of the SDN controllers in the context of the proposed ST system are summarized as follows:

- **VON Creation:** The carrier SDN controller aggregates clients' VON requests, and then embeds the VONs in the physical network by allocating the network resources required by each VON in a centralized way. Once a VON is established, the carrier SDN controller incubates a client SDN controller and the latter is responsible for the management of the newly established VON. The client has full control of the VON via its own SDN controller.
- **Trading Agreement:** To enable spectrum trading between different VONs, all the involved parties must first agree to trade resources between each other. They need to sign an electronic agreement to join the trading community at the beginning, before they can start to trade spectrum resources.
- **VON Resource Utilization Monitoring:** The SDN controller of each VON continues monitoring the spectrum resource utilization to decide whether it has unused spectrum resources or if it lacks spectrum resources.
- **ST Execution:** The ST process is executed in a distributed manner, where each VON client is an equal peer and they communicate with each other directly to decide spectrum trading between themselves. Specifically, a RC broadcasts its spectrum trading request to all the other VON clients who have agreed to join the trading community. Then each client controller that receives the request judges its eligibility to trading according to its own spectrum resource usage. If eligible, it becomes a CC and instructs the RC on the availability of resources for trading. The RC collects the potential trading information from all the CCs and decides a proper set of CCs for actual trading, which then become the corresponding TCs. Finally, the RC instructs all the TCs to provide their unused spectrum resources for it to use.
- **Network Reconfiguration:** Based on the trading agreement made in the previous trading execution step, the SDN controllers of all the TCs make their unused spectrum resources ready to be used by the RC and the SDN controller of the RC uses these newly traded spectrum resources to reconfigure its VON to support more traffic demand. All these reconfigurations are physically supported by the operator's central controller upon the requests of the RC's and the TCs' SDN controllers.
- **Spectrum Resource Release:** There are two situations

triggering the release of spectrum resources obtained from trading. The first situation is that the trading agreement for a certain resource expires. Then this resource must be released following the agreement. The second situation is that the VON that provides spectrum resources may suddenly have a surge of traffic demand and need to take back its own spectrum resources to carry its own traffic. In this case, the RC must release the spectrum resources of the TC upon notification and then reinitiate a new spectrum trading call with CCs to compensate for the released spectrum resources.

- **Trading Record Maintenance:** Each SDN controller of the TCs involved in a trading process creates a trading record to record the amount of spectrum resources that it has contributed to the community. This trading record is maintained by a blockchain, which is described in the next section.

It should be noted that as an operational overhead, the above ST process may cause some network service disruption due to the reconfiguration of the bandwidth of a virtual link involved in trading. However, given the fast reconfiguration speed of the SDN system, this disruption would be trivial compared to the long duration of each trading time slot.

### *D. Blockchain-Assisted Trading Information Maintenance Layer*

The trading data is critical to the operation of the ST system, where a secured data-recording approach is strongly required. In conventional systems, an encrypted database can be applied as one possible way to secure data in a centralized fashion. However, this type of solution still has some potential risks, such as a denial of service (DoS) attack, a database failure, etc. To enhance the data security of the ST system, we propose to use a blockchain-based database to record the trading data for all the involved VON clients. This is similar to many applications such as online voting and smart package tracking, where even though a generic database approach can be also used, a blockchain can provide better security in a distributed implementation.

Figure 3 illustrates trading records formatted in blockchain blocks. Each block consists of a header and a trading record. The header is the hash value of the previous block. The trading record contains the trading information (transactions) of a ST process, which includes: 1) Serial number: A dedicated serial number differentiates a transaction from others. 2) Time slot index: Index of the time slot that the trading record is created for. 3) RC/TC ID: The identifier number (ID) identifies the client that participates in the ST process. 4) Physical link index: Index of a physical link, on which the ST process occurs. 5) Traded FS indexes: Indexes of FSs traded in the ST process. There can be multiple physical links and FSs involved in the trading. Thus, in each transaction, there would be multiple link-FS index pairs.

To maintain the blockchain, a consensus mechanism is required to guarantee its integrity and consistency and to ensure an unambiguous ordering of transactions and blocks. The most widely used protocol for this is proof-of-work (PoW) [12], where a complex mathematical puzzle is required to be solved for the generation of a new block. However, PoW is not efficient in the ST system that needs immediate transaction finality and has high transaction rates [13]. Moreover, a PoW typically has an extremely high power consumption. To address these limitations, we propose a new consensus protocol called *proof-of-contribution* (PoC). In PoC, the creator of a new block is chosen in a deterministic way, based on the amount of the traded spectrum resources. Specifically, the creator is a VON client that provides the largest amount of spectrum resources to the ST community. The selected client generates a new block for the transactions that record the trading information. This block is broadcast to all the VON clients in the system. All VON clients need to confirm the accuracy of the new block by running a verification function before adding it to their own blockchain database. While the main power consumption in most conventional blockchain systems is attributed to solving the PoW, here with PoC, the blockchain-assisted data recording is more time- and energy-efficient since the consensus can be more quickly and easily achieved.

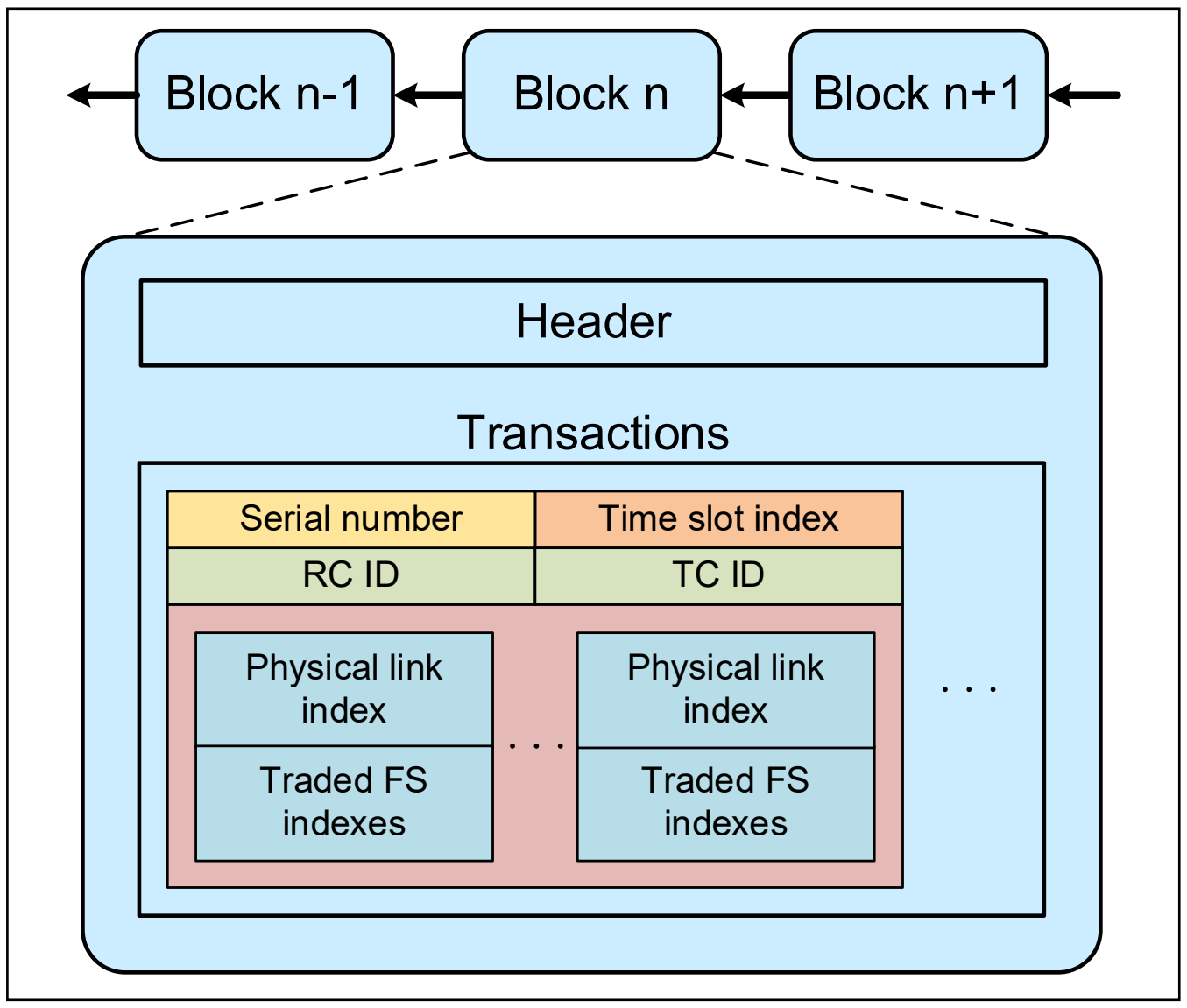

**Figure 3.** Trading record formatted in blockchain blocks.

## IV. Performance Evaluation

This section gives some preliminary results to show the benefits of the proposed ST scheme. We employed the USNET network in Fig. 4 as the test network. The physical distance (in units of km) of a lightpath is employed to decide the modulation format used according to the transparent reach table (see Table 1). We assumed that 50 VONs have been set up and their lifecycles are the same and are uniformly divided into four time slots. The numbers of virtual nodes and virtual links of each VON are randomly generated within the ranges of [$N$/3, 2$N$/3] and [$L$/3, 2$L$/3] respectively, where $N$ and $L$ are the numbers of physical nodes and links, respectively. The capacity $X$ of each virtual link in a VON is the product of the number of FSs assigned and the capacity of each FS. The number of FSs assigned to each virtual link is the same, ranging from 2 to 10 FSs with a 2-FS increasing step. The actual traffic demands on each virtual link in different time slots are randomly generated

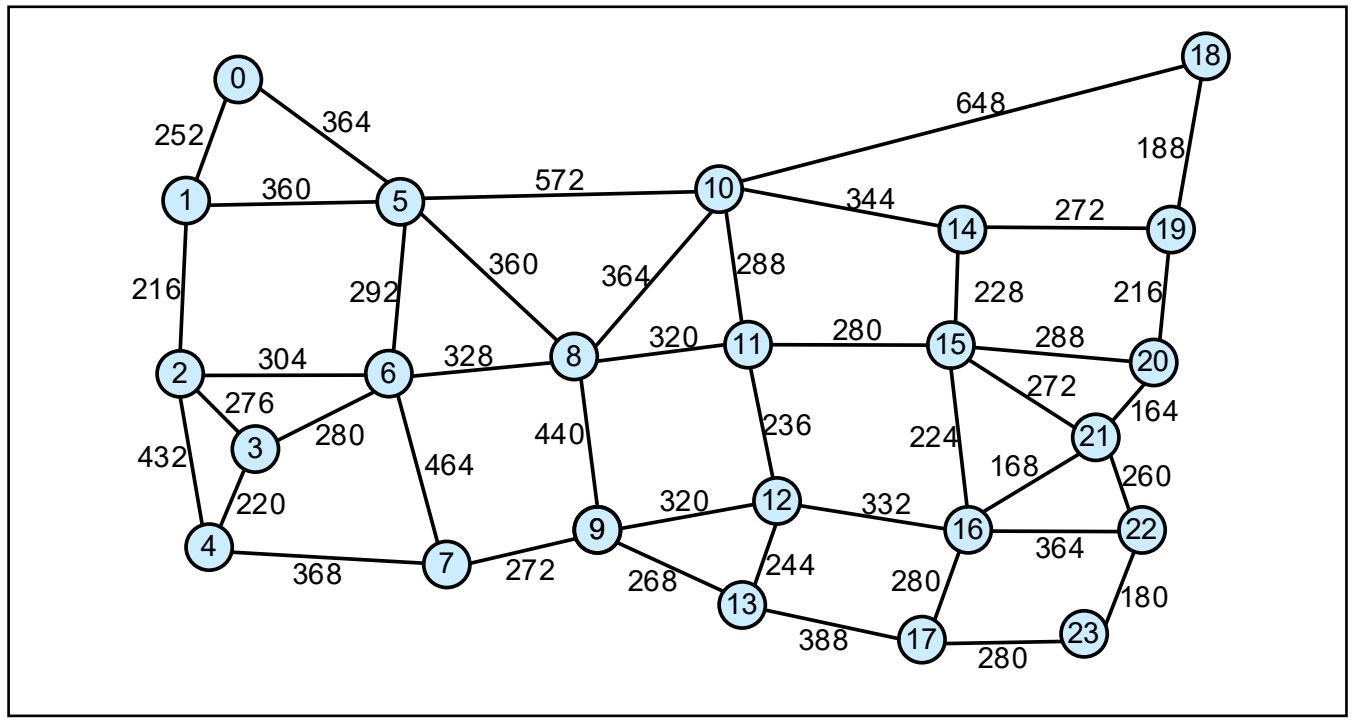

**Figure 4.** Test network.

| Modulation format | SE (bits/symbol) | FS capacity (Gb/s) | Transparent reach (km) |
|---|---|---|---|
| BPSK | 1 | 25 | 4000 |
| QPSK | 2 | 50 | 2000 |
| 8-QAM | 3 | 75 | 1000 |

**Table 1.** FS capacity and transparent reaches of different modulation formats.

within the range of $[10, 2X - 10]$ Gb/s.

Figure 5a compares the amount of overall traffic demand that is actually carried by the VONs for different capacity assigned on each virtual link, where legends "ST" and "non-ST" correspond respectively to the cases with and without spectrum trading. This case study sets the threshold of cumulative credit for ST to be $\mu$=-30. We can see that with the increase of the capacity assigned, the total amount of traffic demand carried by the VONs also increases. Moreover, the case with ST can always carry more traffic demand than the case without ST. With the increase of the capacity assigned, the difference between them increases and reaches more than 21%. This is because a larger capacity assigned corresponds to a larger fluctuation of the actual traffic demand. This triggers more spectrum trading between VONs to improve the overall capacity utilization.

Figure 5b shows the impact of the cumulative credit threshold on the performance improvement achieved by spectrum trading. We can see that with an increasing cumulative credit threshold, the traffic demand established through spectrum trading outperforms the case without spectrum trading by up to 18% with 4 FSs assigned on each virtual link and 21% with 8 FSs assigned on each virtual link. However, this improvement may not keep increasing and demonstrates a saturation when the threshold is too large. This indicates that a small but suitable cumulative credit threshold can not only guarantee the fairness of the ST community but is also adequate to provide an efficient performance.

## V. Open Research Issues

To evoke more research interest and to explore further the benefits of the ST scheme, we next highlight several open issues.

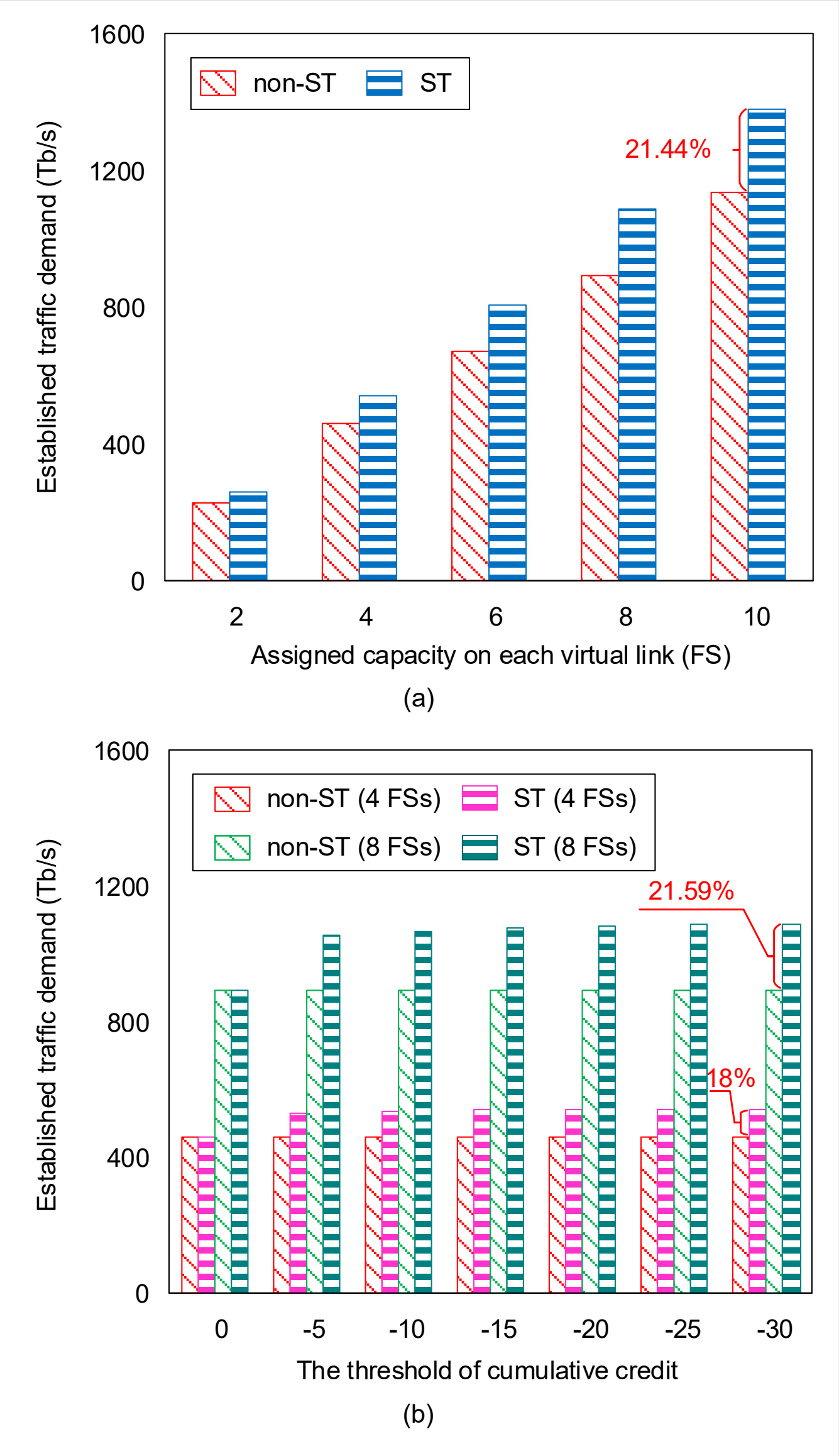

**Figure 5.** Simulation results of the ST scheme: a) established traffic demand vs. assigned capacity on each virtual link ($\mu$=-30); b) established traffic demand vs. threshold of cumulative credit.

### A. *Trading Scope*

The case study in this paper is performed based on the assumption of sub-band VCAT, under which a virtual link can be established by multiple separate (not spectrally neighboring) sub-bands. In practice, some transponders may not support sub-band VCAT. Thus, as a further study, a scenario requiring the spectrum contiguity of a lightpath should be considered. In addition, this study assumes that spectrum trading can be carried out between any VONs that share common physical links. To simplify network operation, we may allow trading only between the virtual links of the same end nodes. This can avoid a complicated signaling process for lightpath reconfiguration, as it would only require tuning the bandwidth at the two end nodes of the virtual links involved in trading.

### *B. Spectrum Trading under Dynamic VON Establishment*

In this study, the ST scheme is evaluated based on a static scenario, where all the VONs are simultaneously embedded in a physical network and have the same service duration. More practically, we may consider the scenario where new VONs' establishment and old VONs' release are also dynamic. More comprehensively, we may also consider an evolutionary case where the traffic demand on each VON increases gradually. Though more complicated, the same ST process can be applied for these scenarios with suitable extensions.

### *C. Spectrum Trading Group*

In the proposed ST scheme, clients can trade their spectrum resources with any other client. However, some clients may not wish to establish a trading relationship with particular clients, e.g., if they are business competitors. In that case, we may need to divide the clients into several spectrum-trading groups (STGs) where the clients within the same STG can trade spectrum resources, but trading between clients not in the same STG would not be allowed. We may also need to coordinate the spectrum resource usage of different STGs since a client may belong to multiple STGs at the same time.

### *D. Unassigned Spectrum*

In the ST scheme, only pre-assigned spectrum resources (owned by clients) can be traded. This may become inefficient when few clients have unused spectrum resources. To make the ST scheme more efficient, we may allow clients to purchase new unassigned spectra from the network operator when all spectra owned by other clients are occupied. In this case, the network operator is also involved in the trading process and an efficient mechanism is required to decide when to request more spectrum resources from the network operator such that the overall costs for the clients are minimized.

## VI. Conclusion

We presented a ST scheme to trade spectrum resources between VONs, which improves the network service qualities of the involved VONs and the overall spectrum utilization of an EON. To achieve flexible network management and secure the integrity of trading records, we also presented a distributed SDN control system and a blockchain-assisted trading information maintenance approach. A performance evaluation was made to verify the efficiency of the proposal, which can achieve up to 21% increase of traffic demand carried by VONs. Finally, we further discussed the open issues of this ST proposal.

## Biographies

**Shifeng Ding** is a Master student with Soochow University in China. His esearch interest focuses on optical networks.

**Kevin X. Pan** is a Product Manager at Twitch.tv (Amazon subsidiary). His expertise lies in understanding cryptocurrency consensus algorithms, often evaluating new consensus algorithms with Bitcoin's famous Proof of Work algorithm.

**Sanjay K. Bose** received the Ph.D. degree from S.U.N.Y. Stony Brook. He is a Professor with IIT Guwahati. He has been working in various areas in the field of computer networks and queuing systems, and has published extensively in the area of optical networks and network routing.

**Qiong Zhang** received the Ph.D. degree from the University of Texas at Dallas. She was an Assistant Professor with Arizona State University. She joined the Fujitsu Laboratories of America in 2008. She was a recipient of the Best Paper Award of GLOBECOM 2005, ONDM 2010, ICC 2011, and NetSoft 2016.

**Gangxiang Shen** is a Distinguished Professor with Soochow University in China. He has co-authored more than 150 technical papers. He was a Guest Editor of two IEEE JSAC special issues and an Associated Editor of IEEE/OSA JOCN. Currently he is an Associate Editor of IEEE Networking Letters, OSN, and PNET. He is an IEEE ComSoc Distinguished Lecturer (2018-2019) and a voting member of IEEE ComSoc Strategic Planning Standing Committee (2018-2019).